# Density Functional Studies Reveal Anomalous Lattice Behavior in Metal Cyanide, $AgC_8N_5$

Baltej Singh[1,2], Mayanak K. Gupta[1], Ranjan Mittal[1,2] and Samrath L. Chaplot[1,2]
[1]Solid State Physics Division, Bhabha Atomic Research Centre, Mumbai, 400085, India
[2]Homi Bhabha National Institute, Anushaktinagar, Mumbai 400094, India

We have investigated anomalous lattice behavior of metal organic framework compound $AgC_8N_5$ on application of pressure and temperature using *ab-initio* density functional theory calculations. The van der Waals dispersion interactions are found to play an important role in structural optimization and stabilization of this compound. Our *ab-initio* calculations show negative linear compressibility (NLC) along the c-axis of the unit cell. The ab-initio lattice dynamics as well as the molecular dynamics simulations show large negative thermal expansion (NTE) along the c-axis. The mechanism of NLC and NTE along the c-axis of the structure is governed by the dynamics of Ag atoms in the a-b plane. The NLC along the c-axis drives the NTE along that direction.

The understanding of atomic level mechanisms responsible for various functional properties [1-8]of materials is very important to improve their performance. By intuition, material should contract (or expand) on application of hydrostatic pressure (or temperature), yet a small number of crystals show opposite behaviour in a few directions. This type of anomalous behaviour with pressure (or temperature) is called negative linear compressibility (or negative thermal expansion) behavior[9]. Materials with a negative linear compressibility (NLC) and negative thermal expansion (NTE) could have interesting technological applications[9-11] in body muscle systems(as actuators), in marine optical telecommunication, components of aviation, sensors, and devices working in high pressure- temperature surroundings.

So far, few studies have been performed to realize the origin of NLC and NTE in crystalline solids[9, 12]. This type of behaviour is predicted in open framework structure with high unit cell volume per atom and low density (Table I). Many cyanide based metal organic flexible framework structure[4, 5, 13-15] like $ZnAu_2(CN)_4$, $M_3Co(CN)_6$, $MAuX_2(CN)_2$ where M=H, Au, Ag, Cu and X=CN, Cl, Br etc. show anomalous NLC and NTE behaviour. It seems that the occurrence of NLC is closely related to the NTE behaviour in these compounds[2].Moreover, there is a very strong link between NLC, NTE and geometry of the material[3, 4, 8, 16].

The NLC and NTE in $ZnAu_2(CN)_4$ arise from anharmonic nature of low energy optic phonon modes involving bending of the -Zn-NC-Au-CN-Zn- linkage[2]. This bending produces the effects of a compressed spring upon heating and an extended spring[17] under hydrostatic pressure in specific spring like topology of $ZnAu_2(CN)_4$. In many cases like MFM-133(M) (M=Zr, Hf), L- tartrate, $M_3Co(CN)_6$ (M=H, Ag, Cu), the anomalous lattice behavior has been observed to arise from deformation mode of wine-rack-like geometries which is contributed from molecular strut compression and angle opening mechanism[3, 5, 6, 9, 13, 14, 18, 19]in anisotropic framework. In $M_3Co(CN)_6$, the intense and quicker response of phonon group velocity along c-axis than in a-b plane upon heating and compression facilitate c-axis to be a carrier of anomalous lattice behavior[20]. In a framework material, NLC can be effectively tuned by varying the inorganic component of the framework without changing the network topology and structure[18]. For 2-D layered framework compounds like $Co(SCN)_2(pyrazine)_2$, the layer sliding mechanism is found to be responsible for the observed NLC behaviour[21]. Framework hinging mechanism leads to an extreme NLC in $InH(BDC)_2$, which is much higher than CN based metal organic framework compounds[22]. Large NLC is observed in frameworks composed of rigid linear ligands and flexible framework angles[2, 4, 23].The NTE behavior along c-axis in linear chain structure like MCN (M=Au, Ag, Cu) is caused by the chain sliding phonon modes and $–C\equiv N–$ bond flipping in the chain[24].

On the basis of current understanding of the mechanisms responsible for NLC, different approaches to design and fabricate new structures with NLC and NTE behaviour are being studied[25],[26].High- pressure and high- temperature X- ray and neutron diffraction techniques are used to experimentally determine the anisotropic linear compressibility and linear thermal expansion coefficients of crystalline materials[3-5, 7-9, 16, 23]. However, ab-initio quantum mechanical calculations are well established to understand the microscopic mechanism governing these phenomena [2, 12, 24, 27-29]. The compressibilities and expansion coefficients as calculated using ab-initio DFT and phonon calculations are found to reproduce fairly well the experimental values. These calculations, from the analysis of eigen vectors, provide the insight about the anharmonic phonon modes responsible for anomalous lattice behaviour in the material [2, 24, 27, 29].

The metal organic framework compound, $AgC_8N_5$ has comparatively lower density[30](Table I) than its family compounds[16, 24] like AgCN and $AgC_4N_3$.This motivated us to study the anomalous lattice behaviour of $AgC_8N_5$ using ab-initio density functional theory. The details of the calculations are given in supplementary information[31].

As per our knowledge, there are no temperature/pressure dependent experimental or theoretical studies reported on this compound regarding its anomalous lattice behavior. The crystal structure (Fig. 1) of $AgC_8N_5$ contains the $C_8N_5$ planer ligand and $AgN_4$ tetrahedral units. Both, the planer and polyhedral units are distorted. The planer sheets of $C_8N_5$ are vertically placed along c-axis with small tilting in the a-b plane. These sheets are well separated along b-axis (distance $> \approx 3$ Å) in a-b plane. Therefore, these sheets must be weakly interacting through weak dispersion interaction to make the stable structure. The structure optimizations done without considering these weak interactions is found to highly overestimate (Table II) the b- lattice parameters. However, when van der Waals interactions are considered between these planers sheets the calculated structure is found to match with the experimental structure[30], within the limitations of GGA (Table II). It seems van der Waals interactions play a very important role in governing the structure stability of $AgC_8N_5$. The weak dispersion interactions acting in a-b plane (especially along b- axis) make the structure flexible in a-b plane as compared to that along c-axis. The structure with presence of van der Waals dispersion interactions is considered for all further calculations.

The elastic constants of $AgC_8N_5$ are derived from the strain−stress relationships obtained from finite distortions of the equilibrium lattice. The elastic constants are used to get the Bulk modulus and elastic compliance matrix, $S=C^{-1}$ (in $10^{-3} GPa^{-1}$ units) as:

$$\begin{pmatrix} 162.95 & -99.12 & -42.81 & 0 & 0 & 0 \\ -99.12 & 143.22 & 7.97 & 0 & 0 & 0 \\ -42.81 & 7.97 & 30.81 & 0 & 0 & 0 \\ 0 & 0 & 0 & 435.16 & 0 & 0 \\ 0 & 0 & 0 & 0 & 202.80 & 0 \\ 0 & 0 & 0 & 0 & 0 & 203.79 \end{pmatrix}$$

For negative compressibility along the crystallographic axes[32], in an orthorhombic crystal, the following inequalities should hold

$$X_a=S_{11} + S_{12} + S_{13}<0,\ X_b=S_{12} + S_{22} + S_{23}<0\ \&$$
$$X_c=S_{13} + S_{23} + S_{33}<0$$

Where $X_i$ (i=a, b, c) are the compressibilities of crystal along various crystallographic axes. It is observed that only the last inequality holds, implying that compound exhibits negative linear compressibility along the c-axis. To quantify this property, the crystal structure is relaxed under application of isotropic pressures and corresponding lattice parameters are calculated. The calculated lattice parameters as a function of pressure are shown in Fig 2. It is observed that the lattice parameters 'a' and 'b' decrease with increasing pressure and show normal behavior. The 'b' lattice parameter shows a larger decrease as compared to 'a' lattice parameter. This arises from the soft nature of van der Waals dispersion interaction acting among the planer sheets of $C_8N_5$ along b-axis. However, the 'c' lattice parameter shows an increase with increasing pressure. This confirms the negative linear compressibility along the c-axis. Overall volume is found to decrease with increase in pressure. Bulk modulus is calculated from the pressure dependence of unit cell volume. The PV equation of state is fitted with the well-known Birch–Murnaghan (2$^{nd}$ ordered) isothermal equation of state to get the value of bulk modulus. The calculated bulk modulus using this approach has the value of 14.5 GPa. This is consistent with that calculated from the elastic compliance (14.48 GPa) matrix. At ambient pressure, the calculated linear compressibilities are found to have the values of $X_a=21.0\times 10^{-3}$, $X_b=52.1 \times 10^{-3}$ and $X_c=-4.0 \times10^{-3} GPa^{-1}$.

We observe that atomic coordinates of Ag atoms show major change on application of pressure. The displacement corresponding to this change is indicated as vectors in Fig 1. The displacement vector shows that the Ag atoms displace along the a-axis of the crystal. Further we found significant displacement of C≡N unit, as a single rigid unit, attached to the corresponding Ag atoms. We found that C≡N units only connected to Ag atoms are displaced in the a-b plane. Therefore, a hinging movement of rigid C≡N units occurs about Ag atoms. On application of pressure, this gives rise to an expansion of the framework along c-axis and contraction in a-b plane.

As pressure increases above 4GPa, the structure becomes unstable and undergoes an unusual change. Around this pressure there are sudden jumps in lattice parameters, total energy and bond lengths of the compound (Fig 2). A sudden decrease in volume gives a signature of high pressure phase transition of the crystal. As indicated above, we found an anomalous change in the Ag atomic coordinate at this pressure. Above 4.0 GPa, $AgN_4$ tetrahedra are found to change the coordination and converts to $AgN_5$. This pressure may correspond to some phase transition of the structure; however, detailed high-pressure diffraction would be needful to identify the resultant structure of new phase.

The calculated structure as a function of pressure shows that the C≡N bond remains unchanged with a value of about 1.16Å. Moreover, various C-C bonds (Fig 4) in -$C_8N_5$- planer units do not show any significant changes (less than 1-2 %) with increase in pressure. The Ag-N bonds of $AgN_4$ tetrahedral units show (Fig 4) a variation of 5-6 %. This change would be due to the flexible nature of Ag-N bonds. There are significant changes (up to 6% of original bond angle) in the C-C≡N bond angle (Fig 4) present on the periphery of $C_8N_5$ structural units. To understand these structural changes and the mechanism for negative and positive linear compressibility along c- axis and in the a-b plane respectively, we have calculated the difference in atomic coordinates of all the atoms corresponding to ambient pressure structure and 4GPa structure.

The primitive unit cell of $AgC_8N_5$ contains 168 atoms and has 504 phonon modes of vibrations. The calculation of complete phonon spectra in the entire Brillouin zone is computationally expensive for such a large system. We have calculated the zone centre phonon spectra in conventional unit cell with 336 atoms using linear response density functional perturbation theory (DFPT). These 1008 phonons at the zone- centre of the conventional unit cell correspond to 504 phonons each at the zone- centre and the (111) zone-boundary point of the Brillion zone of the body centred orthorhombic structure. Leaving 3 acoustic branches, we have calculated the phonon spectrum for 1005 phonon modes. The calculated partial density of states of C, N and Ag atoms show that these atoms contribute in different energy regions (Fig 3) of the spectra. The very high energy peaks around 270 meV in the spectra of C and N is related to the vibrational stretching modes associated with very strong C≡N bonds. The spectra in the 100-200meV range is highly contributed by C atoms and is related to stretching vibrations of strong C-C bonds in the -$C_8N_5$-planer structural units. The lower energy modes in spectra of C and N atoms are associated with bending vibrations of constructing -$C_8N_5$-units. It is interesting to note that the Ag atoms only contribute to the vibrational spectra in very low energy up to 40 meV. Hence, the vibrational modes associated with Ag atoms would be populated at low temperatures and would give rise to interesting structural and dynamical properties. This also confirms, on application of pressure, the dominant role of Ag in giving rise to NLC and phase transition in $AgC_8N_5$.

The stress dependence of phonon energies is used for the calculation of anisotropic Grüneisen parameters of $AgC_8N_5$. An anisotropic stress of 5 kbar is implemented by changing only one of the lattice constants and keeping the others fixed. The calculated mode Grüneisen parameters as a function of phonon energy along different crystal directions are shown in Fig. 4(a). The Grüneisen parameters along a and b axes show normal positive behavior. However, the Grüneisen parameters along c-axis have large negative values. The calculated anisotropic Grüneisen parameters and elastic compliance matrix are used to calculate the anisotropic thermal expansion coefficient. The calculated temperature dependence of anisotropic thermal expansion coefficients and lattice parameters are shown in Fig 4 (c,d). Negative thermal expansion is found along c-axis while positive along a and b axes. The change with temperature along a-axis is more pronounced than that along b-axis. This is contrary to the calculated behavior as a function of pressure. The volume shows normal positive thermal expansion behaviour. The quantitative thermal expansion behavior is obtained from the calculated temperature dependence of linear thermal expansion coefficients (Fig 9). At 300K, the values of linear and volume thermal expansion coefficient are found to be $\alpha_a=105.4\times10^{-6}K^{-1}$, $\alpha_b=32.3\times10^{-6}K^{-1}$, $\alpha_c=-98.1\times10^{-6}K^{-1}$ and $\alpha_V=39.6\times10^{-6}K^{-1}$ respectively. These values are comparable to those reported for the highly anomalous thermal expansion cyanides[3, 7, 16] like $AgC_4N_3$, $Ag_3Co(CN)_6$, $ZnAu_2(CN)_4$ etc. The quasi-harmonic approximations has been proven very good for studying the NTE materials like[2, 12, 24, 28, 29] $LiAlSiO_4$, $V_2O_5$, MCN (M=Ag, Au, Cu), $ZrW_2O_8$, $Ag_3Co(CN)_6$, and $ZnAu_2(CN)_4$.

A significant contribution to thermal expansion behaviour is also reported to arise from the explicit anharmonicity due to finite thermal amplitudes in a few compounds [28, 33-35]. Therefore, we have calculated the temperature dependence of lattice parameters using ab initio molecular dynamical (MD) simulations (Fig 4 (c)). These calculations reproduce the anomalous thermal expansion behavior as calculated from the quasi-harmonic lattice dynamics (LD).

In order to understand the mechanism of anomalous thermal expansion behavior in $AgC_8N_5$, we have calculated the contribution of individual phonon modes to the linear thermal expansion coefficients (Fig 4(b)). It is observed that the low energy phonon modes around 5meV are responsible for the observed negative thermal expansion along c-axis and positive thermal expansion along a and b axes. The displacement patterns of two modes with energy 3.34 meV and 5.32 meV are shown in Fig 5. The low energy phonon modes have dominant contributions from the Ag atoms. These two modes (assuming them as Einstein modes with one degree of freedom each) gives the linear thermal expansion of $\alpha_a=4.3\times10^{-6}K^{-1}$, $\alpha_b=-1.0\times10^{-6}K^{-1}$, $\alpha_c=-3.1\times10^{-6}K^{-1}$ and $\alpha_a=0.3\times10^{-6}K^{-1}$, $\alpha_b=0.4\times10^{-6}K^{-1}$, $\alpha_c=-0.2\times10^{-6}K^{-1}$ at 300K respectively. The mode at 3.34 meV involves the rotation of Ag atoms around the connecting $-C_8N_5-$ structural units in the a-b plane. On the other hand, the mode at 5.32 meV involves displacement of Ag atoms along b- axis. The motion of Ag produces effect of a closing hinge exactly opposite to that observed on application of pressure. This gives rise to contraction along the c-axis and expansion along a and b axes.

In conclusion, our ab-initio DFT calculations reveal large anomalous lattice behaviour in $AgC_8N_5$, which is a metal organic framework material with very low crystal density. Extensive calculations as a function of pressure reveal negative linear compressibility along the c-axis. Moreover, the pressure dependent phonon calculations performed using linear response density functional perturbation theory methods show anomalous thermal expansion behaviour in this compound. The temperature dependence of lattice parameters with (MD) and without (LD) explicit anharmonic effects agrees very well. The NLC and NTE along c-axis of the structure are governed by the dominant dynamics of Ag atoms in a-b plane which give rise to hinge-like mechanism. The compound $AgC_8N_5$ may be very useful for strong armour applications due to its anomalous lattice behaviour.

**Acknowledgements**

S. L. Chaplot would like to thank the Department of Atomic Energy, India for the award of Raja Ramanna Fellowship. The use of ANUPAM super-computing facility at BARC is acknowledged.

TABLE I. The experimental unit cell volume per atom, density and linear thermal expansion coefficients ($\alpha_l$) at 300 K of metal cyanides from literature. The values of $\alpha_l$ ($l$=a, b,c) for $AgC_8N_5$ are obtained from our ab-initio calculations.

| Compound | V/atom ($Å^3$) | Density (g/cm$^3$) | $\alpha_l\times10^{-6}$ K$^{-1}$ | | |
|---|---|---|---|---|---|
| | | | $\alpha_a$ | $\alpha_b$ | $\alpha_c$ |
| AgCN[36] | 13.6 | 4.07 | 66 | 66 | –24 |
| $AgC_4N_3$[16] | 15.5 | 2.63 | -48 | 200 | -54 |
| $AgC_8N_5$ | 16.3 | 2.00 | 105 | 32 | -98 |
| $Ag_3Co(CN)_6$[3] | 19.1 | 2.93 | 132 | 132 | -130 |
| $ZnAu_2(CN)_4$[4] | 19.3 | 4.37 | 37 | 37 | -58 |
| $KMnAg_3(CN)_6$[23] | 19.8 | 2.83 | 61 | 61 | -60 |

TABLE II. The comparison of calculated and experimental lattice parameters of $AgC_8N_5$.

| Optimization Scheme | a (Å) | b (Å) | c (Å) | V (Å$^3$) |
|---|---|---|---|---|
| DFT-GGA (0K) | 12.46 | 15.91 | 32.27 | 6392 |
| DFT-GGA+ vdW (0K) | 12.26 | 13.25 | 32.61 | 5297 |
| Experimental (300K) | 12.43 | 13.62 | 32.30 | 5466 |

FIG 1 (Colour online) (**Top**) The crystal structure of $AgC_8N_5$ containing $AgN_4$ (Blue) tetrahedra connected through C (Red) and N (Green) atoms. (**Bottom**) The calculated displacement pattern giving rise to negative linear compressibility along c-axis of $AgC_8N_5$. The arrow represents the displacement vector for Ag as obtained from the difference in atomic coordinates of all the atoms corresponding to ambient pressure and 4 GPa structure

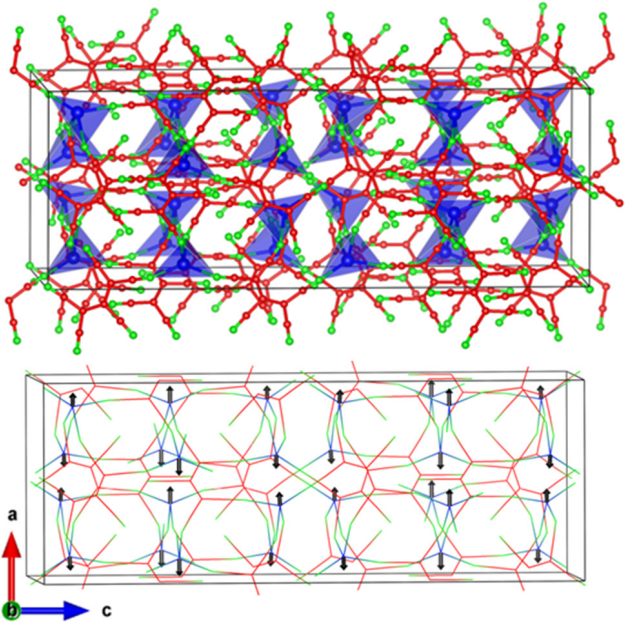

FIG 2 (Colour online) The calculated pressure dependence of lattice parameters (*l*), bond lengths and total energy/atom for $AgC_8N_5$.

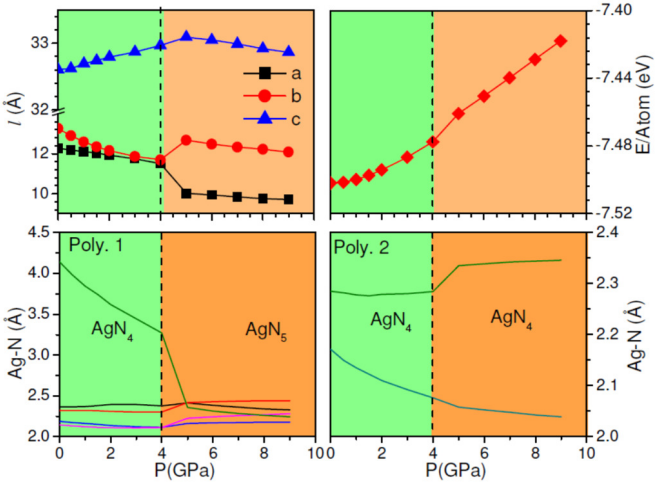

FIG 3 (Colour online) The calculated total and partial phonon density of states for various atoms of $AgC_8N_5$.

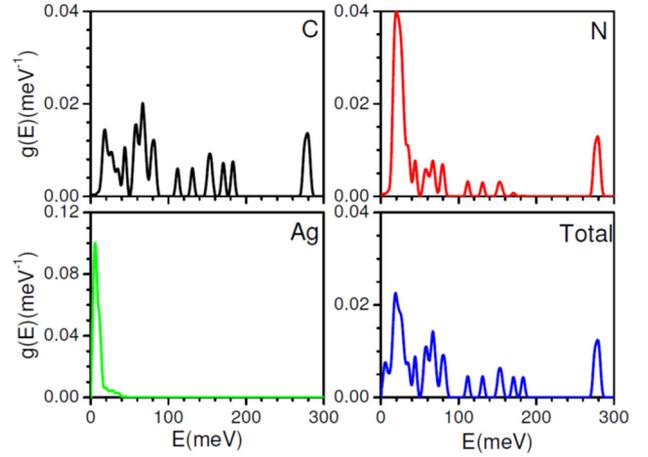

FIG 4 (Colour online) (a) The calculated energy dependence of anisotropic Grüneisen parameters, Γ for $AgC_8N_5$. (b) The contribution of phonon mode of energy E to the linear thermal expansion coefficients. (c, d) The calculated temperature dependence of linear thermal expansion coefficients and lattice parameters of $AgC_8N_5$.

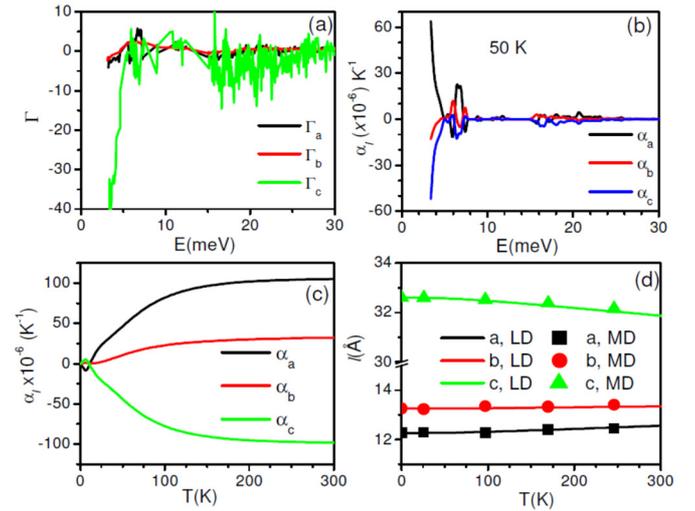

FIG 5 (Colour online) The displacement pattern of optic phonon modes projected in a-b plane of $AgC_8N_5$. The arrow represents the displacement vector for Ag atoms. The displacement vectors for C and N atoms are negligible and are not shown for clarity.

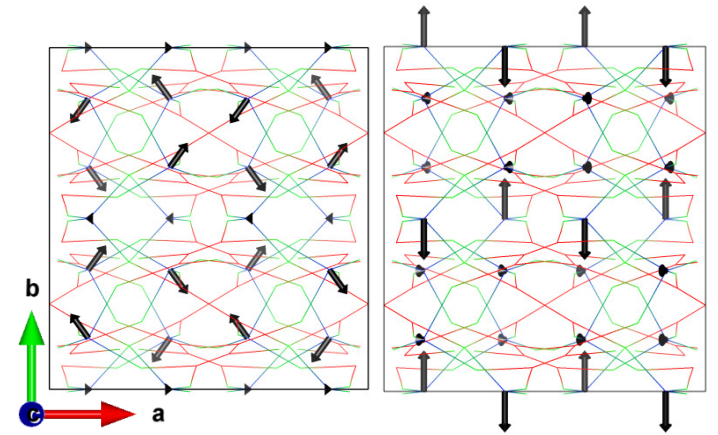

[1] F. Bridges, T. Keiber, P. Juhas, S. J. L. Billinge, L. Sutton, J. Wilde, and G. R. Kowach, Physical Review Letters **112**, 045505 (2014).
[2] M. K. Gupta, B. Singh, R. Mittal, M. Zbiri, A. B. Cairns, A. L. Goodwin, H. Schober, and S. L. Chaplot, Physical Review B **96**, 214303 (2017).
[3] A. L. Goodwin, M. Calleja, M. J. Conterio, M. T. Dove, J. S. O. Evans, D. A. Keen, L. Peters, and M. G. Tucker, Science **319**, 794 (2008).
[4] A. B. Cairns, et al., Nat Mater **12**, 212 (2013).
[5] A. L. Goodwin, D. A. Keen, and M. G. Tucker, Proceedings of the National Academy of Sciences **105**, 18708 (2008).
[6] J. C. Michael, L. G. Andrew, G. T. Matthew, A. K. David, T. D. Martin, P. Lars, and S. O. E. John, Journal of Physics: Condensed Matter **20**, 255225 (2008).
[7] A. L. Goodwin, B. J. Kennedy, and C. J. Kepert, Journal of the American Chemical Society **131**, 6334 (2009).
[8] A. L. Goodwin, D. A. Keen, M. G. Tucker, M. T. Dove, L. Peters, and J. S. O. Evans, Journal of the American Chemical Society **130**, 9660 (2008).
[9] A. B. Cairns and A. L. Goodwin, Physical Chemistry Chemical Physics **17**, 20449 (2015).
[10] N. C. Burtch, J. Heinen, T. D. Bennett, D. Dubbeldam, and M. D. Allendorf, Advanced Materials (2017).
[11] S. M. Mirvakili and I. W. Hunter, Advanced Materials **29** (2017).
[12] R. Mittal, M. K. Gupta, and S. L. Chaplot, Progress in Materials Science **92**, 360 (2018).
[13] A. K. David, T. D. Martin, S. O. E. John, L. G. Andrew, P. Lars, and G. T. Matthew, Journal of Physics: Condensed Matter **22**, 404202 (2010).
[14] A. F. Sapnik, X. Liu, H. L. B. Boström, C. S. Coates, A. R. Overy, E. M. Reynolds, A. Tkatchenko, and A. L. Goodwin, Journal of Solid State Chemistry **258**, 298 (2018).
[15] J. S. Ovens and D. B. Leznoff, Inorganic chemistry **56**, 7332 (2017).
[16] S. A. Hodgson, J. Adamson, S. J. Hunt, M. J. Cliffe, A. B. Cairns, A. L. Thompson, M. G. Tucker, N. P. Funnell, and A. L. Goodwin, Chemical Communications **50**, 5264 (2014).
[17] L. Wang, H. Luo, S. Deng, Y. Sun, and C. Wang, Inorganic chemistry **56**, 15101 (2017).
[18] Y. Yan, A. E. O'Connor, G. Kanthasamy, G. Atkinson, D. R. Allan, A. J. Blake, and M. Schröder, Journal of the American Chemical Society (2018).
[19] H. H.-M. Yeung, R. Kilmurray, C. L. Hobday, S. C. McKellar, A. K. Cheetham, D. R. Allan, and S. A. Moggach, Physical Chemistry Chemical Physics **19**, 3544 (2017).
[20] L. Wang, C. Wang, H. Luo, and Y. Sun, The Journal of Physical Chemistry C **121**, 333 (2016).
[21] Q. Zeng, K. Wang, Y. Qiao, X. Li, and B. Zou, The Journal of Physical Chemistry Letters **8**, 1436 (2017).
[22] Q. Zeng, K. Wang, and B. Zou, Journal of the American Chemical Society **139**, 15648 (2017).
[23] A. B. Cairns, A. L. Thompson, M. G. Tucker, J. Haines, and A. L. Goodwin, Journal of the American Chemical Society **134**, 4454 (2012).
[24] M. K. Gupta, B. Singh, R. Mittal, S. Rols, and S. L. Chaplot, Physical Review B **93**, 134307 (2016).
[25] A. Ghaedizadeh, J. Shen, X. Ren, and Y. M. Xie, Materials & Design **131**, 343 (2017).
[26] K. K. Dudek, D. Attard, R. Caruana-Gauci, K. W. Wojciechowski, and J. N. Grima, Smart Materials and Structures **25**, 025009 (2016).
[27] B. Singh, M. K. Gupta, S. K. Mishra, R. Mittal, P. U. Sastry, S. Rols, and S. L. Chaplot, Physical Chemistry Chemical Physics **19**, 17967 (2017).
[28] M. K. Gupta, R. Mittal, and S. L. Chaplot, Physical Review B **88**, 014303 (2013).
[29] B. Singh, et al., Journal of Applied Physics **121**, 085106 (2017).
[30] L. Jäger, C. Wagner, and W. Hanke, Journal of Molecular Structure **525**, 107 (2000).
[31] See Supplymentary material for computational details, elastic properties and thermal expansion calculations.
[32] C. N. Weng, K. T. Wang, and T. Chen, Advanced Materials Research **33-37**, 807 (2008).
[33] P. Lazar, T. Bučko, and J. Hafner, Physical Review B **92**, 224302 (2015).
[34] G. Ernst, C. Broholm, G. R. Kowach, and A. P. Ramirez, Nature **396**, 147 (1998).
[35] S. L. Chaplot, R. Mittal, and N. Choudhury, *Thermodynamic Properties of Solids: Experiments and Modeling* (John Wiley & Sons, 2010).
[36] S. J. Hibble, G. B. Wood, E. J. Bilbé, A. H. Pohl, M. G. Tucker, A. C. Hannon, and A. M. Chippindale, Zeitschrift für Kristallographie Crystalline Materials **225**, 457 (2010).